\documentclass[aps,prl,superscriptaddress,showpacs,floatfix,twocolumn,amsmath]{revtex4}

\usepackage{graphicx}
\usepackage{hyperref} 

\begin{document}

\title{Local electrical characterization of resonant magnetization motion in a single ferromagnetic sub-micrometer particle in lateral geometry.}

\author{A. Slobodskyy}%

\affiliation{%
Karlsruhe Institute of Technology (KIT)
Light Technology Institute (LTI)
Kaiserstraße 12,
76131 Karlsruhe
Germany and Zentrum f\"{u}r Sonnenenergie- und Wasserstoff-Forschung Baden-W\"{u}rttemberg,
Industriestr. 6,
70565 Stuttgart,
Germany}%
\affiliation{%
Physics of Nanodevices
Zernike Institute of Advanced Materials
University of Groningen
The Netherlands
}%
\author{B. J. van Wees}%
\affiliation{%
Physics of Nanodevices
Zernike Institute of Advanced Materials
University of Groningen
The Netherlands
}%

\date{\today}

\begin{abstract}

In this article a detailed characterization of a  magnetization motion in a single sub-micrometer and multi-terminal ferromagnetic structure in lateral geometry is performed in a GHz regime using direct DC characterization technique. We have shown applicability of the Stoner-Wohlfarth model to the  magnetic nano-structure with large length to with ratio. Applying the model to experimental data we are able to extract relevant magnetization motion parameters and show a correlation between high frequency inductive currents and local magnetization. Additionally, DC voltage generated over the structure at the resonance, with external magnetic field under an angle to the shape anisotropy axis, is explained by the model.
\end{abstract}

\maketitle

Multi-terminal ferromagnetic nanostructures are widely used in spintronics \cite{jedema2001, jedema2002, Lou2007}. Recent developments in the field \cite{costacheapl2006, costacheprl2006, parkin2008} brought enhanced interest to the topic, a specially in combination with ultrafast magnetization dynamics \cite{Mourachkine2008}. Proper inside into magnetization dynamics of an individual nanostructure is a challenging task \cite{thirion2003}. Most of the work in this direction is being done on relatively large structures \cite{soohoo}, arrays of structures \cite{biasi1978}, by interpreting magnetization switching events \cite{parkin2008}, or by dynamic response to the magnetization motion \cite{kiselev2003}. This methods are mostly limited to two terminal measurements. In this article we use multi-terminal ferromagnetic sub-micrometer structures in lateral geometry, in combination with resonant magnetization motion, for detailed quantitative electrical characterization of the sample magnetization. We also analyze the DC voltages generated by the magnetization precession.

\begin{figure}
\centerline{\includegraphics{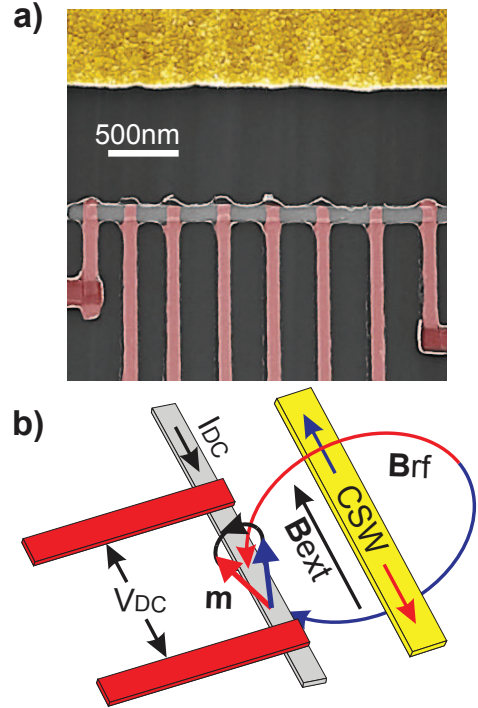}} \caption{\label{sample} a)
SEM image of the device, b) and c) evaporation under an angle fabrication technique, b) perpendicular to the strip and c) along the strip, d) schematic representation of the experiment.}
\end{figure}

The sample is prepared by means of electron beam lithography and lift-of. Scanning electron microscopy (SEM) image of the sample is shown in a Fig.~\ref{sample}(a). The sample consist of a ferromagnetic 3~$\mu$m long and 20~nm thick permalloy (Ni$_{20}$Fe$_{80}$) strip, contacted by eight 50nm thick copper contacts. Width of the strip and the contacts is 100~nm. The strip is placed on 1~$\mu$m distance from a 150~nm thick and 1~$\mu$m wide golden coplanar waveguide. In order to ensure good electrical contacts and uniform magnetization, the permalloy strip and the copper contacts are fabricated by two steps of Electron Beam Evaporation under 70$^{0}$ angle to the sample plane and along corresponding lines, without breaking vacuum.

The measurement technique is depicted in a Fig.~\ref{sample}(b). Coplanar waveguide is used to apply normal to the substrate radio frequency (RF) magnetic field to the permalloy strip. At the resonance, when the RF frequency matches Larmor precession frequency, the field induces Larmor precession of the strip magnetization around the static effective magnetic field ($B_{eff}$). $B_{eff}$ determining Larmor precession frequency is a vector sum of an applied external field ($B_{ex}$) and a constant shape anisotropy field ($B_{an}$) pointing along the strips longest (easy) axis. At the resonance the angle of the magnetization precession is drastically increased  and damped (no precession) otherwise. 

Dependence of s single-domain ferromagnet resistance on relative angle $L$ between magnetization $\textbf{M}$ and current $\textbf{I}$ is called anisotropic magnetoresistance (AMR) \cite{Wegr}, and commonly found to follow the next equation: 

\begin{equation}\label{AMR}
R=R_0+(\Delta R) cos^{2}(L), 
\end{equation}

\noindent where $R_0$ is a resistance when magnetization and current are perpendicular, and $\Delta R$ is a resistance difference between parallel and perpendicular configurations.

In presented experiments outer contacts to the strip are used as DC current probes and the inner contacts as voltage probes. Zero magnetic field resistance in this configuration found to be $R_0 = 121.5~\Omega$, this value is consistent with bulk material properties. By fitting AMR in the case of $B_{ex}$ perpendicular to $B_{an}$ using Eq.~\eqref{AMR}, values of $B_{an} = 0.137~mT$ and $\Delta R = 2.1~\Omega$ are extracted. We use these parameters to extract the average precession angle L at the ferromagnetic resonance using  Eq.~\eqref{AMR}.

\begin{figure}
\centerline{\includegraphics{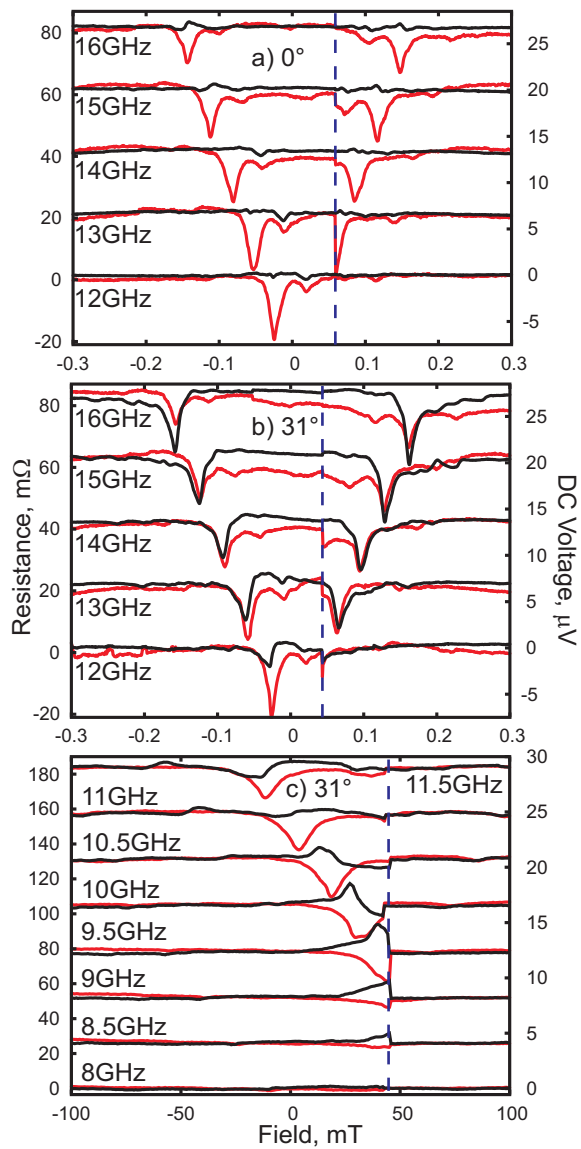}} \caption{\label{data1}
RF frequency dependence of resistance (red curve) and voltage (black curve) versus external magnetic field. a) Field is parallel to the ferromagnetic strip. b) and c) Field is under 31$^{0}$ to the strip easy axis.}
\end{figure}

In Fig.~\ref{data1}(a) RF induced change of the resistance (red line) and change in DC voltage (black line) as a function of $B_{ex}$, applied along the strip are shown. The measurements are done at fixed RF frequency while external magnetic field is scanned from negative to positive direction with amplitude of 300~mT. Magnetization training procedure is used to ensure monodomain state of the magnetization in the strip. Magnetization switching event is shown by a dashed (vertical) line. RF signal is (on, off) modulated at 17~Hz and DC current is alternated between -0.1~mA, 0~mA and +0.1~mA every 10~s. Background voltage extracted from the measurements at opposite current directions is similar to the voltage at zero current. Offset in the resistance and voltage is subtracted, and the curves are shifted for clarity, by 20~$m\Omega$ and 6.7~$\mu V$ respectively. Clear indication of the resonance in the magnetization motion is seen as a deep on the resistance curves, but can not be observed in the voltage measurements.

The measurements at external magnetic field applied under 31$^{0}$ angle to the strip easy axis are shown in Fig.~\ref{data1}(b). Most pronounced difference in this configuration from the previous plot, is that the resonance can be observed as a deep in the voltage measurements. The DC voltage generated at the resonance can be explained by interplay between asymmetry in AMR caused by magnetization precession motion in phase with inductive AC currents trough the ferromagnetic strip. Although, similar results have been observed previously \cite{costacheapl2006}, a systematic analysis over wide range of RF frequencies is done for the first time. Additionally we present a Stoner-Wohlfarth particle model to explain the observed dependence.

In the Fig.~\ref{data1}(c) the 31$^{0}$ angle data is zoomed in the frequency range from 8~GHz till 11.5~GHz and the field range from -100~mT till 100~mT. Resonance in magnetization motion is preserved over zero external magnetic field and defined by the shape anisotropy field. The voltage at the resonance, is changing from a deep for negative external magnetic fields, to a peek for positive external magnetic fields, just before the magnetization switching. This behavior will be explained further.

\begin{figure}
\centerline{\includegraphics{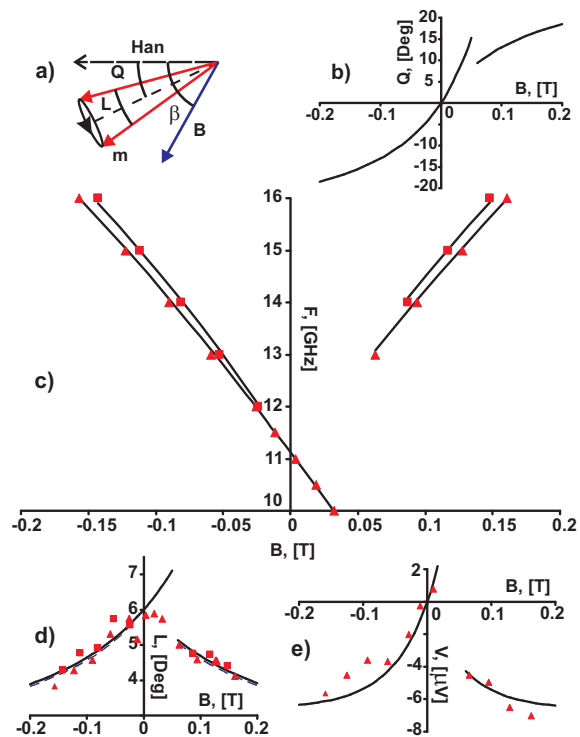}} \caption{\label{fit} External magnetic field dependence of experimental (squares - 0$^{0}$ and triangles - 31$^{0}$) and calculated (lines) parameters. a) Sketch of the magnetization motion.  b) Angle of displacement of the precession axis $Q$ from the shape anisotropy field. c) Average angle of the precession $L$ at the resonance. d) DC voltage generated over the strip at the resonance. e) Resonance RF frequency.
}
\end{figure}

To describe the observed results, when external magnetic field is applied under an angle $\beta$ to the easy axis of the ferromagnetic strip, we calculate expected angle between the axis and the magnetization $Q$, as depicted in the Fig.~\ref{fit}(a). We find $Q$, by energy minimization, as a solution of the following equation: $(H_{an}/2) \sin (2Q) + H_{ex} \sin (Q-\beta)=0$. Solution for 31$^{0}$ tilt is shown in Fig.~\ref{fit}(b). From the figure it can be seen that the precession axis decline as much as $\pm$~20$^0$ at $\pm$~200~mT from the easy axis. Rapid decrease in the angle at about 50~mT is connected with the magnetization switching event.

Using the extracted angles and the shape anisotropy field $B_{an}$ we calculate effective magnetic field $H_{eff}$ as a function of applied external field $B_{ex}$. Than we extract center of each resonance and full width at half maximum (FWHM) by fitting observed resonances in the magnetoresistance curves with Lorentzian \cite{Kittel1953}. RF frequency $F$ of the resonants is shown in Fig.~\ref{fit}(c).

The data below 12~GHz is extracted from Fig.~\ref{data1}(c). The resonance positions in the case of external magnetic fields parallel to the longest strip axis are shown by squares, and the magnetic fields under an angle of 31$^{0}$ to the axis are shown by triangles. Frequency dependence of the resonance position is well fitted with Kitel's equation for small precession angles: $F=\mu_0 \gamma \sqrt{H_{eff}(H_{eff}+M_s)}$, where $\mu_0$ is a permeability of a vacuum, $\gamma$ is the gyromagnetic ratio and $M_s$ is a saturation magnetization.

Despite slight ellipticity expected in the magnetization motion, excellent agreement between fitting and the extracted data is achieved with $\gamma$ = 193.39~GHz and $Ms$ = 969000, the fitting is shown in Fig.~\ref{fit}(c) by lines. The previously extracted shape anisotropy field $B_{an}$=137~mT is used for AMR curve fitting. During magnetization switching 5~mT of the shape anisotropy field are pinned due to sample imperfection.

Taking into account the modulation technique used in our experiment, we find that the measured value of the RF field induced change in the resistance is not sensitive to the angle of the precession axis $Q$, and defined only by magnetization precession angle $L$.

Average precession angle at the resonance $L$ extracted from the data using Eq.~\eqref{AMR} is shown in Fig.~\ref{fit}(d). The data at 31$^{0}$ is shown by triangles and at 0$^{0}$ by squares. Solving transverse RF field $H_{y0}$ driven Landau-Lifschitz-Gilbert equation \cite{gilbert} we find that the precession angle $L$ at a resonance is inversely proportional to the RF frequency $\omega$: $L=A/\lambda\omega$ \cite{guan}, where $A=\mu_{0}^{2}\gamma^{2}M_{s}H_{y0}$ = 6.95 * 10$^8$ and the Gilbert damping parameter $\lambda$ = 0.0104 similar to the value observed previously \cite{gerrits}. Calculated value of $L$ at 31$^{0}$ is shown by continuous line and the value at 0$^{0}$ by dashed line. Although, the measured values are scattered around the calculated lines, smearing separation between angular dependence, it is obvious that the general magnetic field dependence is well preserved.

At positive fields, just before the magnetization switching event the average precession angle $L$ is saturated, contrary to expected strong increase. The saturation could be explained by deviation of the magnetization motion from the Stoner-Wohlfarth particle model.

To determine Gilbert damping parameter we use known \cite{plat} frequency dependence of FWHM  

\[
\Delta H_{FWHM}(\omega) =  \Delta H_{inhom} + \frac{2}{\sqrt{3}} \frac{\lambda}{\gamma^2 M_{s}} \omega ,
 \]

\noindent where $\Delta H_{FWHM}(\omega)$ is a frequency dependent value of FWHM extracted from Lorentzian fit and $\Delta H_{inhom}$ is a frequency independent inhomogeneous broadening due to local magnetization variations. Frequency independent (within experimental error) value of $\Delta H_{FWHM}(\omega) \approx 16~mT \pm 2~mT$ at negative magnetic fields (before the switching) and $\Delta H_{FWHM}(\omega) \approx 21~mT \pm 2~mT$ at positive fields (after the switching) is an indication of dominating contribution of $\Delta H_{inhom}$. Moreover, higher value of the inhomogeneous broadening after the magnetization switching, is consistent with an increase in magnetic inhomogeneity by the pinned magnetization.

In Fig.~\ref{data1} DC voltages generated over the strip shown by black line. Absence of significant voltages at 0$^{0}$ is an indication that the voltages are connected to the magnetization precession symmetry around strip longest axis. In Fig.~\ref{fit}(d) voltages generated at the resonance with the external magnetic field tilt 31$^{0}$ off the strip axis are shown by triangles.

To understand the generated voltages we use already described model for the magnetization motion, taking into account that an inductive current generated trough the sample at the resonance is in phase with the magnetization motion. According to the model the difference in AMR at opposite phases in combination with the inductive current, see Fig~\ref{fit}(a), creates an average DC voltage over the ferromagnet. Measured voltage should, therefore, change the sign when the magnetic field crosses zero, see Fig.~\ref{fit}(b). The voltage changes the sign back to original after $2\pi$ phase shift in the precession motion caused by magnetization switching. Calculated magnetic field dependence of DC voltage, using current of 140~$\mu A$, is shown by lines in Fig.~\ref{fit}(d). The calculations reproduce all the essential features of the measured data.

In summary, we have presented a method of a resonant magnetization motion detection in multi-terminal ferromagnetic sub-micrometer structures. Change in the AMR response at the ferromagnetic resonance is exploited to investigate the magnetic nano-structure. DC resistance response of the structure at the resonance, when the external magnetic field is applied under an angle between the strip easy axis, is explained. DC voltage generated at the resonance under an angle is also well understood, taking into account inductive AC RF currents generated in the ferromagnetic strip.

Using the technique it is now possible to gain new inside on magnetic properties of the ferromagnetic nano-meter scale particles and their differences from a bulk or a layer ferromagnets. Local nature of the measurement technique makes it possible to monitor magnetization motion dynamics in different parts of a single ferromagnetic particle. Investigations in this direction are currently underway.

\begin{acknowledgments}
The authors would like to thank to Prof. Caspar H. van der Wal for useful discussions, as well as NanoNed for financial support.
\end{acknowledgments}

\end{document}